\documentclass[letterpaper ,10pt,journal,draftcls,onecolumn]{IEEEtran}
\usepackage{graphicx}
\usepackage{psfrag}
\usepackage[cmex10]{amsmath}
\usepackage{cite,flushend,amssymb,color,pst-sigsys}

\title{Influence of Doppler Bin Width on GPS Acquisition Probabilities}

\author{Bernhard C. Geiger,~\IEEEmembership{Student Member,~IEEE}, and Christian Vogel,~\IEEEmembership{Senior Member,~IEEE}\thanks{Bernhard C. Geiger is with the Signal Processing and Speech Communication Laboratory, Graz University of Technology, A-8010 Graz, Austria~(email: geiger@tugraz.at).}\thanks{Christian~Vogel is with the Telecommunications Research Center Vienna (FTW),
A-1220 Vienna, Austria~(email: c.vogel@ieee.org).}}

\def \red {black}

\newcommand{\pfab}{P_{fa}}
\newcommand{\pfan}{\overline{P}_{fa}}
\newcommand{\pdet}{P_{det}}

\newcommand{\pfa}[1]{{P_{fa}}\left(\beta\right)}

\newcommand{\pdetl}[1]{P_{det}(\beta,#1)}
\newcommand{\pdetln}[1]{\overline{P}_{det}(\beta,#1)}

\newcommand{\pFA}{P_{FA}(\beta)}
\newcommand{\pDET}{P_{DET}(\beta)}
\newcommand{\pDETc}{P_{DET}(\beta)}

\renewcommand{\exp}[1]{e^{#1}}

\newcommand{\fd}{f_D}
\newcommand{\hfd}{f_{\hat{k}}}
\newcommand{\Dfd}{\Delta f_{\hat{k}}}
\newcommand{\Dmax}{f_{D,\max}}
\newcommand{\DW}{W}
\newcommand{\Stet}{\Sigma\theta_{\hat{k}}}
\newcommand{\ts}{T_s}
\newcommand{\fs}{f_s}
\newcommand{\fc}{f_{IF}}

\newcommand{\ci}{C}

\newcommand{\tper}{T_{per}}
\newcommand{\tetc}{\theta_{IF}}
\newcommand{\tetd}{\theta_D}
\newcommand{\htetd}{\theta_{\hat{k}}}
\newcommand{\Dtet}{\Delta\theta_{\hat{k}}}
\newcommand{\decmet}{X[\hat{m},\hat{k}]}
\newcommand{\fu}{\overline{f}_{l}}
\newcommand{\fl}{\underline{f}_{l}}
\newcommand{\Si}{\mathrm{Si}}

\newcommand{\rif}{r_{IF}}
\newcommand{\rb}{r_{B}}
\newcommand{\rwoc}{r}

\newcommand{\yi}{y}

\newcommand{\noise}{\eta}
\newcommand{\nnoise}{n}
\newcommand{\noisevar}{\sigma_\noise^2}
\newcommand{\nnoisevar}{\frac{\sigma_\nnoise^2}{2}}
\newcommand{\nnoisevart}{\sigma_\nnoise^2}

\newcommand{\corf}[1]{R_{#1}}
\newcommand{\corfs}[1]{R^2_{#1}}

\DeclareMathOperator{\EXP}{\mathbb{E}}
\newcommand{\expec}[1]{\EXP(#1)}
\newcommand{\prob}[1]{\mathrm{Prob}\left(#1\right)}
\newcommand{\marcum}[2]{\mathrm{Q}_1\left(#1,#2\right)}

\newcommand{\sinc}{\mathrm{sinc}}

\newcommand{\code}{c}

\newcommand{\numD}{K}

\newcommand{\corrcode}{\code[n-\hat{m}]}

\newcommand{\ncode}{{N_{C}}}

\newcommand{\unifdist}[2]{\mathcal{U}\left(#1,#2\right)}
\begin{document}

\maketitle

\begin{abstract}
Acquisition is a search in two continuous dimensions, where the digital algorithms require a partitioning of the search space into cells. Depending on the partitioning of the Doppler frequency domain, more than one cell might contain significant signal energy. We present an expression for the expected values of the cells' energies to analyze the impact of the Doppler bin width on detection and false alarm probabilities.
\end{abstract}


\section{Introduction}
\label{sec:intro}
In Global Navigation Satellite Systems (GNSS), such as GPS, every satellite is transmitting a particular pseudo-random noise (PRN) code, which is known at the receiver. Satellites are acquired by correlating the received signal with local code signals and comparing the results against a threshold. In practice, the local replica of the transmitted code signal differs from the received code signal by a code phase shift (i.e., time lag) and a Doppler shift. Both have to be determined simultaneously in a two-dimensional search. The results of this search, which is usually called acquisition, are required for presetting subsequent stages of the GNSS receiver.

For this two-dimensional search, the continuous time-frequency uncertainty region is divided into cells, each corresponding to a particular Doppler frequency and a particular code phase. Typically, the number of considered code phases is predetermined by the sampling rate and optional decimation/interpolation methods, whereas the width and the number of Doppler bins is \textcolor{\red}{mainly limited by the pull-in range of subsequent signal processing stages~\cite{Grewal_GPSIntertialNavigation}}. 

In many civil GNSS receivers exploiting the GPS L1 C/A code~\cite{GPS_ICD}, the integration period of the correlator is set to the code period of 1 ms. The corresponding Doppler bin widths range from 500 to 667~Hz (see~\cite{Borre_SDRGPS,Tsui_GPSReceivers,Kaplan_GPS,Ward_GPSAcquisition} as well as~\cite{O'Driscoll_SDRAcquisition} and the references therein). Unsurprisingly, the choice of the Doppler bin width strongly influences the acquisition performance: Not only is the Doppler bin width inversely proportional to the number of cells to be searched, it also strongly influences the probability of signal detection. Aside from that, the probability of (false) detection at a Doppler bin adjacent to the correct one increases for small Doppler bins.

Non-coherent acquisition methods take the squared magnitude of the correlation coefficients as a decision metric to overcome unknown carrier phases and possible data modulation. For these methods, the computation of the receiver operating characteristics is a well-investigated field of research. The literature provides detection and false alarm probabilities for single cells~\cite{Kaplan_GPS}, a serial search over all cells with threshold comparison~\cite{Polydoros_SerialSearchP1,Polydoros_SerialSearchP2}, a maximum search~\cite{Cheng_AcquisitionDopplerShift,Iinatti_MaxSelection} and combinations thereof~\cite{Corazza_MAX-TCAcquisition}. In~\cite{Borio_ImpactAcquisitionStrategy}, a comparison of the abovementioned techniques is provided for an L1 GPS receiver. Detection probabilities for an L5 GPS receiver with different algorithms combining data and pilot signals are considered in~\cite{Borio_CombiningTechniquesAcquisition}, from which the comprehensive signal model was largely adopted in this work.  In~\cite{Borio_ImpactAcquisitionStrategy}, both the number of Doppler bins and side lobes resulting from adjacent Doppler bins are considered; the latter are only obtained by means of simulations. Results for the effect of residual Doppler shifts on the acquisition performance have been presented in~\cite{Borio_GalileoBOC,Borio_PhD,Mathis_FFTAcquisition,O'Driscoll_PerfFFT}. In~\cite{Mathis_FFTAcquisition} and~\cite{O'Driscoll_PerfFFT} also the effect of adjacent Doppler bins containing significant energy was analyzed, but detection performance was evaluated only numerically. All these works are lacking an analysis of the influence of the Doppler bin width on detection performance in terms of closed-form expressions.

In this work, we fill this gap by deriving expressions for cell detection probabilities as a function of the Doppler bin width. These cell probabilities are then used to compute global detection and false alarm probabilities, which further depend on the number of Doppler bins. With the help of this theoretical framework, a proper analysis and, maybe even more importantly, a performance-oriented design of GNSS acquisition stages is possible. Moreover, while the focus of this work is on GPS receivers, the results can be applied to other CDMA systems affected by large Doppler shifts as well. Our contribution is thus threefold:
\begin{enumerate}
 \item We introduce an approximation of the cell detection probability as a function of the Doppler bin width by replacing the non-centrality parameter within a bin by its expected value.
 \item Assuming that more than one cell in the two-dimensional search region contains signal energy, we derive an expression for the global detection probability of a serial search.
 \item Finally, we combine the previous two results to evaluate the influence of the Doppler bin width  on the receiver operating characteristics. We show that under certain conditions for a given fixed search range and code domain resolution, small Doppler bins outperform larger ones, contradicting results derived under the prevailing assumption of a single signal cell.
\end{enumerate}
The choice of the Doppler bin width will affect not only the detection and the false alarm probabilities, but also the mean acquisition time and the computational complexity of the acquisition process. The final design decision will take all these effects into account, by balancing them among competing design goals. While this work does not consider these other effects, it intends to provide quantifyable results about the receiver operating characteristics, therefore putting design decisions on a solid mathematical ground.

In order to keep the analysis tractable and to obtain closed-form expressions, we base our work on the following assumptions:
\begin{enumerate}
 \item We assume that the true Doppler frequency is uniformly distributed over the searchable Doppler range. Since in many cases an estimate of the Doppler frequency is available from which the search can be started~\cite{Tsui_GPSReceivers}, this assumption underestimates the performance of the receiver and thus acts as a lower bound on the corresponding probabilities.
 \item We further assume statistical independence of the noise portions from different cells. The validity of this assumption depends on how the decision metric for each cell is obtained, but was shown to be quite good for low oversampling in the code domain~\cite{Borio_ImpactAcquisitionStrategy}, as long as different Doppler bins are evaluated using different signal recordings.
 \item We also assume that only a single cell in each Doppler bin contains signal energy. While this can be assured using averaging correlation~\cite{Starzyk_AveragingCorrelation,Soudan_AC,Fantino_AcquisitionSW} (which would also remove statistical dependencies among noise portions), this assumption does not hold for oversampling in the code domain in general. However, as this work is on the influence of the Doppler bin width, an analysis of the impact of the oversampling factor is left for future work.
 \item Although a serial search with threshold comparison is not state-of-the-art anymore, it yields closed-form solutions without requiring additional knowledge about design decisions. By contrast, an analysis of the hybrid search, which bases its decision on the maximum-energy cell of a subset of the search range, would require specifying how these subsets were formed. In addition to that, it has been shown in~\cite{Borio_ImpactAcquisitionStrategy} that the serial search provides a lower bound on the receiver operating characteristics of state-of-the-art search methods.
 \item Finally, we neglect the effects of data modulation on the detection performance of the receiver. For the envisaged scenario -- GPS L1 C/A code, short integration period -- this assumption is quite reasonable since the probability of a data bit sign transition is sufficiently small. A joint analysis of the effect of Doppler bin width and bit transitions is subject to future investigations, expecially since the (expected) attenuation due to bit transitions seems to depend on the Doppler bin width (cf.~\cite[eq.~(3.15)]{O'Driscoll_PhD}).
\end{enumerate}

These assumptions and simplifications made throughout the following analysis interfere to some extent with practical consideration, in the sense that they only hold for a subset of real-world GPS receivers. We claim, however, that such a limitation is necessary: First, these assumptions do not only make the closed-form solutions for the global detection probability possible, but also allow simple numerical validation. Second, by making these simplifications, the effect of Doppler bin widths becomes more prominent while keeping all other effects fixed. This is vital for getting an intuitive understanding of the effect of bin widths, not only from looking at the simulation results, but also from looking at the mathematical equations.

Future work will analyze the join of different aspects affecting detection performance (such as data modulation, statistical dependence of noise samples, oversampling, etc.) as well the effect of the Doppler bin width on computational complexity and mean acquisition time.

The remainder of this article is organized as follows: In Section~\ref{sec:sigmod} the signal model is introduced, while Section~\ref{sec:acquisitionSystem} gives a detailed analysis of the acquisition process. The main contribution of this work is concentrated in Sections~\ref{sec:propabilities} and~\ref{sec:DopplerInfluence}: The former is devoted to deriving global detection and false alarm probabilities for generalized cell probabilities, while in the latter the influence of the Doppler bin width on cell detection probabilities is discussed. The analytic results are finally verified by extensive simulations in Section~\ref{sec:simres}.

\section{Signal Model}
\label{sec:sigmod}
After front-end filtering, downconversion to the intermediate frequency (IF), and sampling, the signal received from a single satellite can be represented as~\cite{Borio_ImpactAcquisitionStrategy}
\begin{equation}
  \rif[n] = \sqrt{2\ci}\; \yi[n] \cos \Bigl[\left(\tetc+\tetd\right)n-\vartheta \Bigr] + \noise[n] \label{eq:sigmod}
\end{equation}
where $\ci$ is the signal power, $\tetc = \frac{2\pi \fc}{\fs}$ and $\tetd = \frac{2\pi \fd}{\fs}$ are the sampled angular equivalents of the intermediate and Doppler frequencies $\fc$ and $\fd$, respectively, and $\vartheta$ is a phase shift introduced by the transmission and the noncoherent downconversion to IF. If the ideal front-end filter has a single-sided bandwidth equal to $\frac{\fs}{2}$ with $\fs$ being the sampling frequency, the noise signal $\noise[n]$ is assumed to be Gaussian\footnote{If a digital GPS receiver is used, in addition to sampling also an amplitude quantization will be performed, leading to non-Gaussian noise which, as suggested by the Central Limit Theorem, becomes approximately Gaussian after the summation in Fig.~\ref{fig:acquisition}. In this work, however, we do not consider quantization effects and refer the interested reader to, e.g.,~\cite{Curran_Quantization}.} with variance~\cite[pp.~556, Prop.~25.15.2]{Lapidoth_DigitalCommunication}
\begin{equation}
 \noisevar = \frac{N_0 \fs}{2}
\end{equation}
and with an autocorrelation function~\cite{Borio_CombiningTechniquesAcquisition}
\begin{equation}
  \corf{\noise,\noise}[m] = \expec{\noise[n]\noise[n-m]}= \noisevar \delta[m]. 
\end{equation}
In these equations, $\frac{N_0}{2}$ is the two-sided noise power spectral density. The carrier is modulated by
\begin{equation}
 \yi [n] = d[n] \code[n]
\end{equation}
where $d[n]$ is the data message and $\code[n]$ is the binary PRN code, i.e., $\code[n]=\pm 1$. For the sake of simplicity, this work only considers the GPS L1 C/A code with a code period of $\tper=1$~ms and $\ncode=1023$ chips per code period~\cite{Kaplan_GPS}. Furthermore, it is assumed that no data is modulated on the PRN code, i.e., $d[n]=1$. For the scenario envisaged within this work (strong satellite signal, integration period equaling the code period) this assumption is unproblematic since the 20~ms duration of a data bit is significantly larger than the code period. As we verified in a separate set of simulations, the influence of data bit transitions is negligible. This is also in accordance with~\cite{O'Driscoll_PhD}, where the effects of data bit transitions on the non-centrality parameter were analyzed. For modern codes, where the code period is equal to the bit duration (e.g., the GPS L2 CM code) and for weak signal conditions requiring larger integration periods, the effects of bit transitions can be mitigated using non-coherent integration~\cite{Qaisar_CrossCorrelationPerformance} or aided acquisition~\cite{Akos_LowPowerGNSS}.\\

\section{Acquisition System}
\label{sec:acquisitionSystem}
\begin{figure}[t]
 \centering
 \footnotesize
 \psfrag{rif}[cl][cl]{$\rif[n]$}
\psfrag{dec}[cl][cl]{$\decmet$}
\psfrag{dec2}[cl][cl]{$|\decmet|^2$}
 \psfrag{rb}[cl][cl]{$\rb[n]$}
 \psfrag{r}[cl][cl]{$\rwoc[n]$}
 \psfrag{code}[cl][cl]{$\corrcode$}
 \psfrag{sum}[cl][cl]{$\frac{1}{N} \sum \limits_{n=0}^{N-1}$}
 \psfrag{expif}[cl][cl]{$e^{\jmath\left(\htetd+\tetc\right)n}$}
\psfrag{abs}[cl][cl]{$|\cdot|^2$}
 \includegraphics[width=0.47\textwidth]{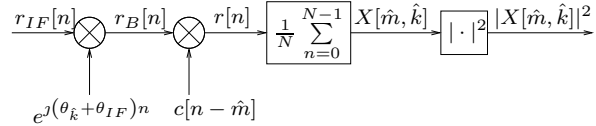}
 \caption{Acquisition of a signal with unknown Doppler frequency and code phase.}
 \label{fig:acquisition}
\end{figure}

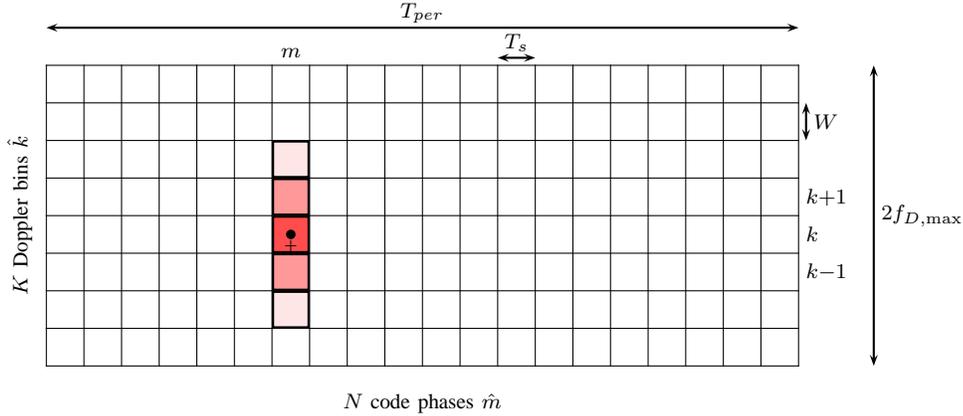
\begin{figure*}[ht]
 \centering
	 \footnotesize
\begin{pspicture}[showgrid=false](-1,-0.5)(12,4.5)
	\psframe[fillcolor=red!70,fillstyle=solid](3,1.5)(3.5,2)
	\psframe[fillcolor=red!40,fillstyle=solid](3,1)(3.5,1.5)\psframe[fillcolor=red!40,fillstyle=solid](3,2)(3.5,2.5)
	\psframe[fillcolor=red!10,fillstyle=solid](3,0.5)(3.5,1)\psframe[fillcolor=red!10,fillstyle=solid](3,2.5)(3.5,3)
  \psgrid[unit=0.5cm,gridwidth=0.4pt,subgridwidth=0.4pt,subgriddiv=1,subgridcolor=black,gridlabels=0](0,-2)(20,8)(0,0)
	\rput[Bc](5,-0.5){$N$ code phases $\hat{m}$}
	\rput[B]{90}(-0.25,2){$K$ Doppler bins $\hat{k}$}
  \rput[B](3.25,4.1){$m$} \rput[l](10.1,1.75){$k$}
  \rput[l](10.1,1.25){$k{-}1$} \rput[l](10.1,2.25){$k{+}1$}
  \psline{<->}(10.1,3)(10.1,3.5) \rput[l](10.2,3.25){$\DW$}
  \psline{<->}(6,4.1)(6.5,4.1) \rput(6.25,4.3){$\ts$}
  \psdots[dotstyle=+](3.25,1.6) \psdots(3.25,1.75)
  \psline{<->}(11,0)(11,4) \rput[l](11.1,2){$2\Dmax$}
  \psline{<->}(0,4.5)(10,4.5) \rput[c](5,4.7){$\tper$}
\end{pspicture}
 \caption{Partitioning of the continuous time-frequency uncertainty region. The correct code phase is denoted by $m$, the correct Doppler bin by $k$. The plus-sign indicates the true Doppler frequency $\tetd$, the dot shows the Doppler estimate $\theta_k$ minimizing the residual Doppler difference $\Dtet$. Note that adjacent Doppler bins can contain significant signal energy (indicated by different shading; cf. Sections~\ref{sec:propabilities} and~\ref{sec:DopplerInfluence}).}
 \label{fig:partitioning}
\end{figure*}

As stated in the introduction, the continuous time-frequency uncertainty region has to be partitioned into cells to make acquisition tractable. Let us, for the remainder of this work, assume that the partitioning of the frequency domain is uniform, and that each of the $K$ resulting Doppler bins has width $\DW$, as shown in Fig.~\ref{fig:partitioning}. Given a maximum expected Doppler frequency\footnote{During the initial acquisition, $\Dmax$ also includes the unknown oscillator frequency bias, which can be of the order of a few kHz and, thus, artificially extends the Doppler search range.} $\pm\Dmax$, the number of bins, $\numD$, is given as
\begin{equation}
 \numD = \left\lceil\frac{2\Dmax}{\DW}\right\rceil. \label{eq:NumBins}
\end{equation}

In the acquisition process illustrated in Fig.~\ref{fig:acquisition}, the received signal $\rif[n]$ is first downconverted using an expected Doppler frequency $\htetd$. The obtained signal $\rb[n]$ can be described, utilizing~\eqref{eq:sigmod}, by
\begin{IEEEeqnarray}{RCL}
 \rb[n] &=& \rif[n] \exp{\jmath (\htetd+\tetc)n}\\
&=& \sqrt{\frac{\ci}{2}} \yi[n] \left( \exp{\jmath (2\tetc+\Stet)n-\jmath\vartheta} +\exp{\jmath\Dtet n+\jmath\vartheta}\right)\notag \\&&  +{}\:\tilde{\noise}[n]\label{eq:rb}
\end{IEEEeqnarray}
with $\Dtet=\htetd-\tetd$ and $\Stet=\htetd+\tetd$. The noise signal $\tilde{\noise}[n]$ is a zero-mean circular-symmetric complex Gaussian (ZMCSCG) signal with variances $\frac{\noisevar}{2}$ for real and imaginary parts. After downconversion the signal is multiplied with the spreading code using an expected code phase $\hat{m}$. Thus,
\begin{IEEEeqnarray}{RCL}
 \rwoc[n] &=& \rb[n] \corrcode\label{eq:r}
\end{IEEEeqnarray}
where $\corrcode$ is the code $\code[n]$ circularly shifted by $\hat{m}$. The decision metric $\decmet$ is obtained by averaging the signal over one code period\footnote{Coherent integration over more than one code period is not directly discussed in this work. However, the analysis can be extended to coherent integration over a (small) integer multiple of one code period by adjusting the carrier-over-noise spectral density ratio $\frac{C}{N_0}$, as long as the effects of data modulation can be assumed to remain negligible~\cite{O'Driscoll_PhD}.} $\tper$,
\begin{equation}
  \decmet = \frac{1}{N} \sum_{n=0}^{N-1} \rwoc[n]
\end{equation}
where the number of samples within one code period is given by
\begin{equation}
 N=\tper\fs \label{eq:NTF}.
\end{equation}
In this operation, the high-frequency terms $\exp{\jmath (2\tetc+\Stet)n-\jmath\vartheta}$ in~\eqref{eq:rb} and~\eqref{eq:r} vanish. 
Following the reasoning in~\cite{Parkinson_GPS}, the influences of time lags and Doppler frequencies can be separated on average. 
We therefore get for the decision metric:
\begin{equation}
  \decmet = \exp{\jmath \Dtet\frac{N-1}{2} +\jmath\vartheta} \frac{\sin\left(\frac{\Dtet}{2}N\right)}{N\sin\left(\frac{\Dtet}{2}\right)} \sqrt{\frac{\ci}{2}}\corf{\yi,\code}[\hat{m}] + \nnoise[\hat{m}]
\end{equation}
where $\corf{\yi,\code}[\hat{m}]$ is the correlation function between $\yi[n]$ and the local code $\code[n]$ evaluated at lag $\hat{m}$ and where the Dirichlet kernel results from summing over $e^{\jmath\Dtet n}$. The noise signal $\nnoise[\hat{m}]$ is the average of $N$ independent ZMCSCG samples, thus the variances of the real and imaginary parts reduce with~\eqref{eq:NTF} to
\begin{equation}
  \nnoisevar = \frac{\noisevar}{2N} = \frac{N_0 \fs}{4N} = \frac{N_0}{4 \tper}.
\end{equation}

The decision is finally based on the squared magnitude of the decision metric $\decmet$, 
\begin{equation}
 |\decmet|^2 = \Re\{\decmet \}^2 + \Im\{\decmet \}^2
\end{equation}
which follows for given $\hat{m}$ and $\hat{k}$ a non-central $\chi^2$-distribution with two degrees of freedom and the non-centrality parameter
\begin{IEEEeqnarray}{RCL}
L_{\hat{m},\hat{k}} 
  &=& 2\tper \frac{\ci}{N_0}\frac{\sin^2\left(\frac{\Dtet}{2}N\right)}{N^2\sin^2\left(\frac{\Dtet}{2}\right)}\corfs{\yi,\code}[\hat{m}]. 
\end{IEEEeqnarray}
Note that the squared means have to be normalized by the corresponding variances, since the $\chi^2$-distribution is defined as the sum of squares of Gaussian random variables with unit variance~\cite[pp.~940]{Abramowitz_Handbook}. With
\begin{equation}
 \Dtet = 2 \pi \frac{\Dfd}{\fs} = 2 \pi \Dfd \frac{\tper}{N}
\end{equation}
and by approximating the Dirichlet kernel by a $\sinc$ kernel we obtain
\begin{equation}
L_{\hat{m},\hat{k}} = 2\tper \frac{\ci}{N_0}\sinc^2\left(\Dfd\tper\right)\corfs{\yi,\code}[\hat{m}].\label{eq:noncentral}
\end{equation}
where $\sinc(x)=\frac{\sin(\pi x)}{\pi x}$. Loosely spoken, the non-centrality parameter is a measure of the ratio between the signal energy and the noise energy within a cell, and thus plays a central role for computing detection probabilities.
                                                                                                                                                                                                                                                                                                                                                                                                                                                                                                                                                                                                                                                                                                                                                                                                                                     
Note that the local spreading code $\code[n]$ has to be upsampled to the sampling rate $\fs$ prior to correlation. Depending on the implementation of the correlation (matched filter, parallel code phase search~\cite{Borre_SDRGPS}, etc.) samples of the decision metric adjacent in the code phase domain are not necessarily statistically independent. This dependence can be removed by applying averaging correlation~\cite{Soudan_AC}, a method which further reduces the computational complexity of the correlation process significantly~\cite{Starzyk_AveragingCorrelation,Fantino_AcquisitionSW}. Note that also in the Doppler domain statistical dependencies may occur: If, for example, all cells are evaluated from the same set of $N$ consecutive input samples (as in some acquisition algorithms employing the FFT), the noise portions of adjacent cells in the Doppler direction may not be statistically independent. For the sake of simplicity, and with the statements about averaging correlation above, however, we assume that the noise portions in different cells are independent, both in the Doppler and in the code phase direction.


\section{Detection Probabilities}
\label{sec:propabilities}
For evaluating the performance of a receiver, the receiver operating characteristics (ROCs) are a key metric. For GPS acquisition, generally two different ROCs can be taken into account: The cell ROC plots the cell detection probability against the cell false alarm probability, and the global ROC plots the global detection probability against the global false alarm probability. For a receiver characterization, the latter one is of main interest, and we thus will focus on global ROCs in the forthcoming sections.

We define a \emph{global detection} as the event that the cell selected by the employed search strategy is the correct cell, i.e., the one with the correct code phase $\hat{m}=m$ and with the correct Doppler index $\hat{k}=k$, where
\begin{equation}
 k=\arg\min_{\tilde{k}} \left\{|\tetd-\theta_{\tilde{k}}| \right\}.
\end{equation}
As will be seen later, the global detection probability depends on cell detection and cell false alarm probabilities. If a cell is chosen in the absence of a signal, we will call this event a \emph{global false alarm}. If the search algorithm chooses no cell at all or the wrong cell in the presence of a signal, neither a false alarm nor a detection occurs. While our definition of a global detection may seem highly restrictive, we will later generalize the definition to accepting also Doppler bins $\hat{k}$ other than the correct one. This generalization does make sense, e.g., in cases where the signal loops in subsequent signal processing stages have sufficiently large pull-in ranges.

We limit ourselves to search strategies employing threshold comparison, i.e., a cell detection or cell false alarm is triggered whenever the decision metric $|\decmet|^2$ for this cell exceeds a certain threshold $\beta$. A decision based on the ratio between the largest and the second largest value of $|\decmet|^2$ of a subset of cells was suggested by~\cite{Jaewoo_(Patent)RatioBasedAcquisition,Jaewoo_RatioImplementation} after the introduction of this ratio as a reliability measure in~\cite{Akos_LowPowerGNSS}. However, the performance of this method has been analyzed just recently~\cite{Geiger_Acquisition}, and it was shown that a decision based on threshold comparison outperforms the ratio detector~\cite{Geiger_Comparison}.

On one hand, as it can be seen from~\eqref{eq:noncentral}, the non-centrality parameter $L_{\hat{m},\hat{k}}$ of the $\chi^2$-distribution is maximized for the correct code phase $m$ and the correct Doppler bin $k$. On the other hand, whenever the desired satellite PRN code sequence is not contained in the received signal, whenever the difference between the actual and the estimated Doppler frequencies is too large, or whenever the code phase is not correct ($\hat{m}\neq m$), the non-centrality parameter $L_{\hat{m},\hat{k}}\approx 0$, neglecting side lobe and cross-correlation levels. Assuming $L_{\hat{m},\hat{k}}=0$ immediately translates to the fact that the $\chi^2$-distribution changes from a non-central to a central distribution.

In this section, we will derive general relations between the cell and the global detection and false alarm probabilities. In particular, we consider not only the number $K$ of Doppler bins, but also take into account that cells with correct code phases, but wrong Doppler indices, may have a non-centrality parameter $L_{m,\hat{k}}>0$.

\subsection{Cell Detection Probabilities}

Let us define the \emph{cell false alarm} as the decision metric $|\decmet|^2$ exceeding a certain threshold $\beta$ given that $L_{\hat{m},\hat{k}} = 0$. Then, the cell false alarm probability becomes~\cite{Iinatti_MaxSelection}
\begin{equation}
 \pfab(\beta) = \prob{|\decmet|^2>\beta\; |\; L_{\hat{m},\hat{k}}= 0}= e^{-\frac{\beta}{\nnoisevart}}.
\end{equation}
Conversely, whenever the decision metric exceeds the threshold for a non-central $\chi^2$-distribution ($L_{\hat{m},\hat{k}}>0$), we will call this event a \emph{cell detection}. Thus, the cell detection probability is
\begin{IEEEeqnarray}{RCL}
 \pdetl{L_{\hat{m},\hat{k}}} &=& \prob{|\decmet|^2>\beta\; |\; L_{\hat{m},\hat{k}} > 0}\notag\\
 &=& \marcum{\sqrt{L_{\hat{m},\hat{k}}}}{\sqrt{\frac{2\beta}{\nnoisevart}}} \label{eq:pdetcell}
\end{IEEEeqnarray}
where $\marcum{\cdot}{\cdot}$ is the Marcum Q-function~\cite{Marcum_PdetRadar,Shnidman_CalculationPdetMarcumQ}. Note that in fact $\pfab(\beta)$ is a special case of $\pdetl{L_{\hat{m},\hat{k}}}$ for $L_{\hat{m},\hat{k}}=0$.

\subsection{Global Detection Probabilities -- Naive Assumption}

Let us now assume that the acquisition is implemented as a serial search over the two-dimensional uncertainty region comprised of $NK$ cells, and the search is stopped when $|\decmet|^2$ exceeds the threshold $\beta$ for the first time. Furthermore, it is assumed that there is only one cell containing significant signal energy, i.e., there is only one ($\hat{m}$,~$\hat{k}$)-pair for which the decision metric $|\decmet|^2$ is non-centrally $\chi^2$-distributed -- namely the pair ($m$,~$k$). 
Following~\cite{Borio_ImpactAcquisitionStrategy}, the global false alarm probability $\pFA$ therefore calculates to
\begin{equation}
 \pFA = 1-\left(1-\pfab(\beta)\right)^{NK} \label{eq:PFA_Borio},
\end{equation}
whereas the global detection probability $\pDET$ can be calculated as
\begin{equation}
 \pDET = \frac{1}{NK} \frac{1-\left(1-\pfab(\beta)\right)^{NK}}{\pfab(\beta)}\pdetl{L_{m,k}}. \label{PDET_Borio}
\end{equation}
For very small values of $\pfab(\beta)$ the above equations can be approximated by $\pFA\approx NK\pfab(\beta)$ and $\pDET\approx\pdet(\beta)$~\cite{Borio_ImpactAcquisitionStrategy}. Consequently, for an increasing number $K$ of Doppler bins (i.e., for smaller bin widths $\DW$) the detection performance degrades. As we will see in Sections~\ref{sec:DopplerInfluence} and~\ref{sec:simres}, the same may not necessarily hold for the refined model.

\subsection{Global Detection Probabilities -- Refined Model}
\label{ssec:adjacent}

The assumption of a single signal cell (i.e., a cell for which $L_{\hat{m},\hat{k}}>0$) is clearly strong, since the side lobe and cross-correlation levels of the correlation function $\corfs{\yi,\code}[\hat{m}]$, a non-zero width of the correlation main lobe for $N>N_c$, and signal energy in Doppler bins adjacent to the correct one will lead to $L_{\hat{m},\hat{k}}>0$ for more than one cell. Effects of side lobes and cross-correlations in the code phase search can be neglected in medium SNR levels or mitigated by appropriate threshold settings, and the effects due to the correlation main lobe can be reduced by means of averaging correlation~\cite{Starzyk_AveragingCorrelation,Fantino_AcquisitionSW,Soudan_AC}. Signal energy in adjacent Doppler bins ($k\pm 1$, $k\pm 2$,\dots), however, not only affects global detection probabilities, but also depends on the width of the Doppler bins. A proper analysis of this influence will be presented in the following.

We now assume that $L_{m,k\pm 1}$, $L_{m,k\pm 2}$,\dots\ are all positive while $L_{\hat{m},\hat{k}}=0$ for all $\hat{m}\neq m$ and all $\hat{k}$. In the absence of the desired PRN code let $L_{\hat{m},\hat{k}}=0$ for all cells. In other words, each Doppler bin $\hat{k}$ contains one signal cell at the correct code phase $m$ (see Fig.~\ref{fig:partitioning}) with an expected non-centrality parameter $L_{m,\hat{k}}$. Again, acquisition is assumed to take place as a serial search with threshold comparison.

We will now generalize the definition of a global detection for stopping the serial search at a Doppler bin in a sufficient adjacency of the correct bin, i.e., at a bin $\hat{k}\in\{k-M,\dots,k+M\}$. The integer $M$ denoting the successful detection range may depend, for example, on the Doppler bin width $\DW$ and the pull-in range of a frequency-locked loop of a subsequent tracking stage. 

Since there is now more than a single signal cell, the direction of the serial search has an influence on the global detection probability. In other words, searching all code phases for each Doppler bin and searching all Doppler bins for each code phase leads to different performance results. Note that both methods can be efficiently implemented using the FFT: In this case, the former option is called parallel code phase search, while the latter is often referred to as parallel frequency search~\cite{Borre_SDRGPS}. Following the reasoning in~\cite{Borio_ImpactAcquisitionStrategy}, the probability of detection for a search over all code phases for each Doppler bin can be calculated as
\begin{multline}
 \pDETc = \frac{1}{KN} \frac{1-\pfan^N(\beta)}{\pfab(\beta)}\sum_{q=-M}^M \pdetl{L_{m,k+q}} \\
   \times{} \left[\sum_{n=n'}^{K'-1} \pfan^{n(N-1)}(\beta) \prod_{l=1}^n \pdetln{L_{m,k-l+q}} \right] \label{eq:pcpsM}
\end{multline}
where $K'=\min\{K,K+q\}$, $n'=\max\{0,q\}$, $\pfan(\beta) = 1-\pfab(\beta)$, and $\pdetln{L_{m,\hat{k}}} = 1-\pdetl{L_{m,\hat{k}}}$. The probability of detection for a search over all Doppler bins for each code phase calculates to
\begin{multline}
  \pDETc =\frac{1}{KN}\frac{1-\pfan^{KN}(\beta)}{1-\pfan^K(\beta)}\times{}\\
\sum_{q=-M}^M \pdetl{L_{m,k+q}}\left(\sum_{n=n'}^{K'-1}\prod_{l=1}^{n}\pdetln{L_{m,k-l+q}}\right).\\ \label{eq:pfsM}
\end{multline}
An outline of the derivation of these results can be found in Appendix~\ref{app:derivation}.

Although theoretically different, for small cell false alarm probabilities $\pfab(\beta)$ we can write $\pfan^M \approx 1-M\pfab(\beta)\approx 1$ and thus obtain as an approximation for both search directions:
\begin{multline}
 \pDETc = \frac{1}{K}\sum_{q=-M}^M \pdetl{L_{m,k+q}}\\\times \left(\sum_{n=n'}^{K'-1}\prod_{l=1}^{n}\pdetln{L_{m,k-l+q}}\right). \label{eq:approxPdM} 
\end{multline}
Depending on the acquisition strategy, this approximation can be shown to hold independently of the SNR by setting the threshold $\beta$ to obtain a constant false alarm rate~\cite{Kaplan_GPS} satisfying $KN\pfab(\beta)\ll 1$. The threshold can be efficiently determined by continuously measuring the noise floor~\cite{Geiger_Comparison}.


It is worth mentioning that the global false alarm probability is the same for the naive and the refined model, i.e., it is always given by~\eqref{eq:PFA_Borio}. Further, when the naive assumption holds, i.e., $L_{\hat{m},\hat{k}}=0$ for $\hat{m}\neq m$ and $\hat{k}\neq k$, both~\eqref{eq:pcpsM} and~\eqref{eq:pfsM} reduce to~\eqref{PDET_Borio}, while the same does not hold for the approximation~\eqref{eq:approxPdM}. Finally, both the naive and the refined model assume that the correct cell is uniformly distributed over the two-dimensional search space. While this assumption can be justified for the code phase domain, depending on the approximate time and position of the receiver, an estimated Doppler frequency can be computed, from which the search should be initiated~\cite{Tsui_GPSReceivers,Grewal_GPSIntertialNavigation}. Since the Doppler frequency is likely close to its estimate if the estimation process was successful, the uniformity assumption provides a lower bound on the detection performance of the receiver.

\section{Influence of Doppler Bin Width on the Decision Metric}
\label{sec:DopplerInfluence}
\begin{figure}[t]
 \centering
 \includegraphics[width=0.47\textwidth]{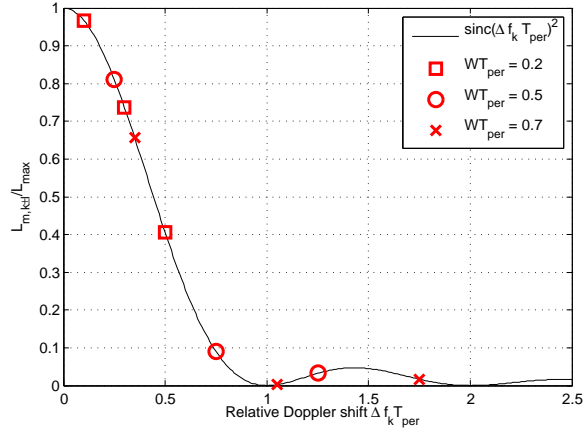}
 \caption{Squared sinc function. Markers indicate Doppler bin boundaries for designated (relative) bin widths $\DW\tper$.}
 \label{fig:sinc}
\end{figure}

We have defined the non-centrality parameter of the $\chi^2$-distribution modeling the decision metric $|\decmet|^2$ in~\eqref{eq:noncentral}, which shows that it is proportional to the squared $\sinc$ of the Doppler difference $\Dfd=\hfd-\fd$, as well as to the squared correlation function $\corfs{\yi,\code}[\hat{m}]$. Using proper decimation methods~\cite{Starzyk_AveragingCorrelation,Fantino_AcquisitionSW,Soudan_AC} and neglecting side lobes and cross-correlation levels we assume that there exists only one code phase $m$ for which the correlation function is non-zero, and that for this phase we have $\corfs{\yi,\code}[m]=1$. In the literature (e.g.,~\cite{Borio_CombiningTechniquesAcquisition,Cheng_AcquisitionDopplerShift}), however, even the squared $\sinc$ is often approximated by unity for the correct Doppler bin, and by zero for all other bins such that
\begin{IEEEeqnarray}{RCL}
 L_{m,k} &=& L_{\max} = 2\tper \frac{\ci}{N_0}\notag\\
 \forall \hat{k}\neq k{:}\ \ L_{m,\hat{k}} &=& 0.
\end{IEEEeqnarray}
As Fig.~\ref{fig:sinc} shows, this simplification is overly optimistic: For small Doppler bins, such as $\DW\tper=0.2$, the Doppler bin adjacent to the correct one (the area between the two leftmost $\square$-markers) has a significant non-centrality parameter $L_{m,k\pm1}$. Moreover, for large Doppler bins (e.g., $\DW\tper=0.7$) the non-centrality parameter $L_{m,k}$ of the correct Doppler bin might be as low as indicated by the leftmost $\times$-marker.
This section is thus devoted to a more in-depth analysis of the influence of Doppler bin widths, which will lead to a more realistic characterization of the acquisition performance.

As stated in Section~\ref{sec:acquisitionSystem} (see~Fig.~\ref{fig:partitioning}), in the acquisition process the whole Doppler domain is divided into bins of equal width $\DW$ which are searched in a serial fashion. The center frequency of a bin represents the Doppler estimate $\hfd =\htetd\frac{\fs}{2\pi}$ with which the received signal is demodulated. Assuming a uniform distribution of the true Doppler frequency $\fd$ we obtain a uniform distribution of the residual Dopper difference in the $k\pm l$-th bin~\cite{Mathis_FFTAcquisition,Borio_GalileoBOC}:
%
\begin{equation}
 \Delta f_{k\pm l} \sim \unifdist{(2l-1)\frac{\DW}{2}}{(2l+1)\frac{\DW}{2}}\label{eq:fAdjacent}
\end{equation}
with $l\in\mathbb{N}_0$. Since now $\Dfd$ is a random variable (RV), also the non-centrality parameter $L$ becomes an RV with probability density function (PDF) $f_{L}(\lambda)$. The detection probability thus calculates to
\begin{equation}
  \pdetl{L} = \int_{\beta}^{\infty} \int_{-\infty}^{\infty}  f_{Y|L}(y|\lambda) f_L(\lambda) d\lambda dy \label{eq:accurate}
\end{equation}
where $f_{Y|L}(y|\lambda)$ is the PDF of a non-central $\chi^2$-distribution with non-centrality parameter $L=\lambda$. This expression is difficult to compute since the PDF of $L$ is not readily available (in~\cite{Mathis_FFTAcquisition}, e.g., only an approximation based on a polynomial expansion has been derived) and also the resulting integrals might not have convenient closed-form solutions. Instead, assuming that the Doppler bins are sufficiently small, it is possible to approximate the PDF of $Y$ to depend linearily on $L$ and we get, as shown in Appendix~\ref{app:linear},
\begin{IEEEeqnarray}{RCL}
  \pdetl{L}  &\approx& \int_{\beta}^{\infty} f_{Y|L}(y,\expec{L}) dy\label{eq:approximate}
\end{IEEEeqnarray}
where $\expec{L}$ is the expected value of $L$.
Exploiting the method for computing the expected value of a function of a random variable from~\cite[pp.~142]{Papoulis_Probability} and using the uniform distribution~\eqref{eq:fAdjacent}, it is straightforward to compute\footnote{For the sake of simplicity we abuse notation and, from now on, write $L_{m,\hat{k}}$ instead of $\expec{L_{m,\hat{k}}}$.} $L_{m,\hat{k}}=L_{m,k\pm l}$
\begin{IEEEeqnarray}{RCL}
  L_{m,k\pm l} 
    &=& 2\tper \frac{1}{\DW}\frac{\ci}{N_0} \int_{(2l-1)\frac{\DW}{2}}^{(2l+1)\frac{\DW}{2}} \sinc^2\left(\Dfd\tper\right)  d(\Dfd).\notag\\
\end{IEEEeqnarray}
Substituting $x=\Dfd\tper$ we get $dx=\tper d(\Dfd)$ and
\begin{IEEEeqnarray}{RCL}
  L_{m,k\pm l} &=&  \frac{2}{\DW}\frac{\ci}{N_0} \int_{\fl\tper}^{\fu\tper} \sinc^2\left(x\right) dx.
\end{IEEEeqnarray}
where $\fl=(2l-1)\frac{\DW}{2}$ and  $\fu=(2l+1)\frac{\DW}{2}$ denote the lower and upper boundaries of the $k\pm l$-th Doppler bin. We obtain
\begin{IEEEeqnarray}{RCL}
  L_{m,k\pm l} 
&=&\frac{2}{\DW}\frac{\ci}{N_0}\frac{1}{\pi} \Big[\Si(2\pi\fu\tper)-\Si(2\pi\fl\tper)\notag\\
&&{}+\frac{\sin^2(\pi\fl\tper)}{\pi\fl\tper}-\frac{\sin^2(\pi\fu\tper)}{\pi\fu\tper} \Big] \label{eq:meanLk}
\end{IEEEeqnarray}
where $\Si(\cdot)$ is the sine integral~\cite[pp.~231]{Abramowitz_Handbook}. For $\underline{f}_0=-\frac{\DW}{2}$ and $\overline{f}_0=\frac{\DW}{2}$ (i.e., for the bin $k$ containing the correct Doppler frequency) this yields
\begin{IEEEeqnarray}{RCL}
 L_{m,k} = \frac{2}{\DW}\frac{\ci}{N_0}\frac{1}{\pi} \left[2 \Si(\pi\DW\tper)-\frac{4\sin^2(\pi\frac{\DW}{2}\tper)}{\pi\DW\tper} \right].\label{eq:meanL0}
\end{IEEEeqnarray}
By the properties of the sine integral 
the difference in~\eqref{eq:meanLk} will contribute significantly whenever at least one of the arguments falls within $[-\pi,\pi]$, i.e., for low values of $l=|\hat{k}-k|$. Thus, one can expect that Doppler bins close to the correct bin have, on average, large non-centrality parameters, while large offsets lead to small non-centrality parameters. This is in line with intuition, which suggests that the correct Doppler bin contains most of the energy. Moreover, in both~\eqref{eq:meanLk} and~\eqref{eq:meanL0} not only the absolute Doppler bin width $\DW$, but also the integration period $\tper$ has an influence on the non-centrality parameter. As a consequence, we will analyze the effect of the \emph{relative} Doppler bin width $\DW\tper$. Since for small values of $\DW\tper$ more bin boundaries may fall in the interval $[-\pi,\pi]$, the correct Doppler bin $k$ on average contains more energy if the bin is small. For large relative bin widths the correct Doppler frequency may be far from the bin center, compared to the former case.

While all this reasoning suggests that smaller Doppler bins are preferable, another aspect has to be taken into account: Given a fixed Doppler search range of $\pm \Dmax$, according to~\eqref{eq:NumBins}, $K$ bins have to be searched. Since $K$ is inversely proportional to the bin width $\DW$, smaller widths lead to a higher number of bins, which increases the probability of false alarms according to~\eqref{eq:PFA_Borio} and decreases the probability of detection (see~\eqref{PDET_Borio} and~\eqref{eq:approxPdM}). Moreover, in cases where $\DW\tper$ is small, bins adjacent to the correct Doppler bin may contain significant signal energy (large $L_{m,k\pm1}$, $L_{m,k\pm2}$, etc.), which can -- depending on the definition of a successful detection -- trigger a false detection and thus degrade receiver performance. As a consequence, depending on the threshold, the integration period $\tper$, and the SNR, there will be an optimal Doppler bin width $\DW$ maximizing the global detection probability $\pDET$.


\section{Simulations and Results}
\label{sec:simres}
To verify the analytic results a series of simulations was conducted. To this end, a set of satellite signals was generated. For simplicity, it was assumed that just a single satellite (GPS L1 C/A PRN code 1, $\ncode=1023$, $\tper=1$~ms) was visible with random Doppler frequency $\fd$ and code phase $m$ (see below). The carrier-over-noise spectral density ratio $\frac{C}{N_0}$ was set to 40~dBHz unless stated otherwise. While this ratio represents a rather strong signal, weaker signal conditions would typically require a longer coherent integration time $\tper$ to keep detection performance acceptable. On the one hand, an increased integration period would increase the \emph{effective} SNR which, assuming the effects of data modulation remain negligible, allows the application of our results (cf.~\cite[Table~5.10]{Tsui_GPSReceivers}). On the other hand, an increased integration period either degrades detection performance by increasing the relative Doppler bin width $\DW\tper$ or leads to a longer acquisition time by decreasing the absolute Doppler bin width $\DW$.

After sampling the signal with a high sampling frequency, it was assumed that prior to correlating over one code period, the signal was decimated to the code chipping rate by means of averaging correlation (cf.~\cite{Starzyk_AveragingCorrelation,Fantino_AcquisitionSW}). Thus, $N=\ncode=1023$. This simplification does not affect the validity of the analysis, since with this method the statistical properties of the cells do not change~\cite{Soudan_AC}. It was assumed, however, that during decimation the correct code phase $m$ corresponding to $\corf{\yi,\code}[m]=1$ is preserved. Aside from that, the correct code phase was uniformly distributed on the set of possible code phases.

For each realized satellite signal, the Doppler frequency was drawn according to a uniform distribution over the whole Doppler range, i.e., $\fd\sim\unifdist{-\Dmax}{\Dmax}$, where $\Dmax=5000$~Hz~\cite{Grewal_GPSIntertialNavigation}. To make both simulation and analytic comparison tractable, only two Doppler bins adjacent to the correct bin contained signal energy, i.e., $L_{m,k\pm l}=0$ for $l=3, 4, \dots$, which leads to the plot depicted in Fig.~\ref{fig:partitioning} and to
\begin{equation}
 \forall l=3, 4, \dots\ \ \pdetl{L_{m,k\pm l}} = \pfa{\beta}
\end{equation}
This simplification holds well for relative bin widths $\DW\tper$ greater than 0.3, as shown in Fig.~\ref{fig:sinc}. The Doppler bin widths were varied within $\DW\in\{200,500,700,1000\}$ Hz corresponding to the relative widths $\{0.2, 0.5, 0.7, 1\}$. Recall further that the number of Doppler bins $K$ is determined by $\Dmax$ and $\DW$ (cf.~\eqref{eq:NumBins}).
The signals were correlated with PRN codes 1 and 5 for the detection and false alarm probabilities, respectively. For each Doppler bin width $\DW$ a set of $10^5$ correlations was performed. 

A serial search is implemented: Starting from the first Doppler bin, all possible code phases are searched sequentially until either the threshold is crossed or until the whole Doppler bin is searched. Then, the next Doppler bin is taken into consideration. If the first threshold crossing occurs at the correct code phase in the correct or in one of its $2M$ adjacent Doppler bins, the signal is assumed to be detected, while any threshold crossing in the absence of a signal triggers a false alarm. For the analytic results for the global detection probability, the accurate expression~\eqref{eq:pcpsM} was used.

\subsection{Influence of Doppler Bin Width on Cell Probabilities}\label{sec:cellProbs}

\begin{figure}[t]
 \centering
 \includegraphics[width=0.47\textwidth]{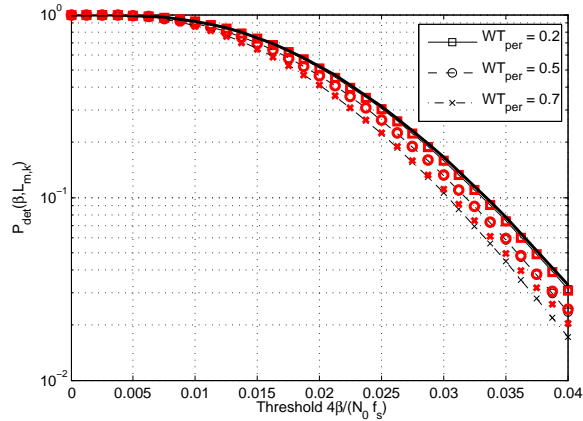}
 \caption{Cell detection probability $\pdetl{L_{m,k}}$ for different Doppler bin widths. Simulated (bold markers) and analytic (lines) results are shown for the correct Doppler bin. The thick solid line indicates $\pdetl{L_{\max}}$.}
 \label{fig:pdet_corr}
\end{figure}

\begin{figure}[t]
 \centering
 \includegraphics[width=0.47\textwidth]{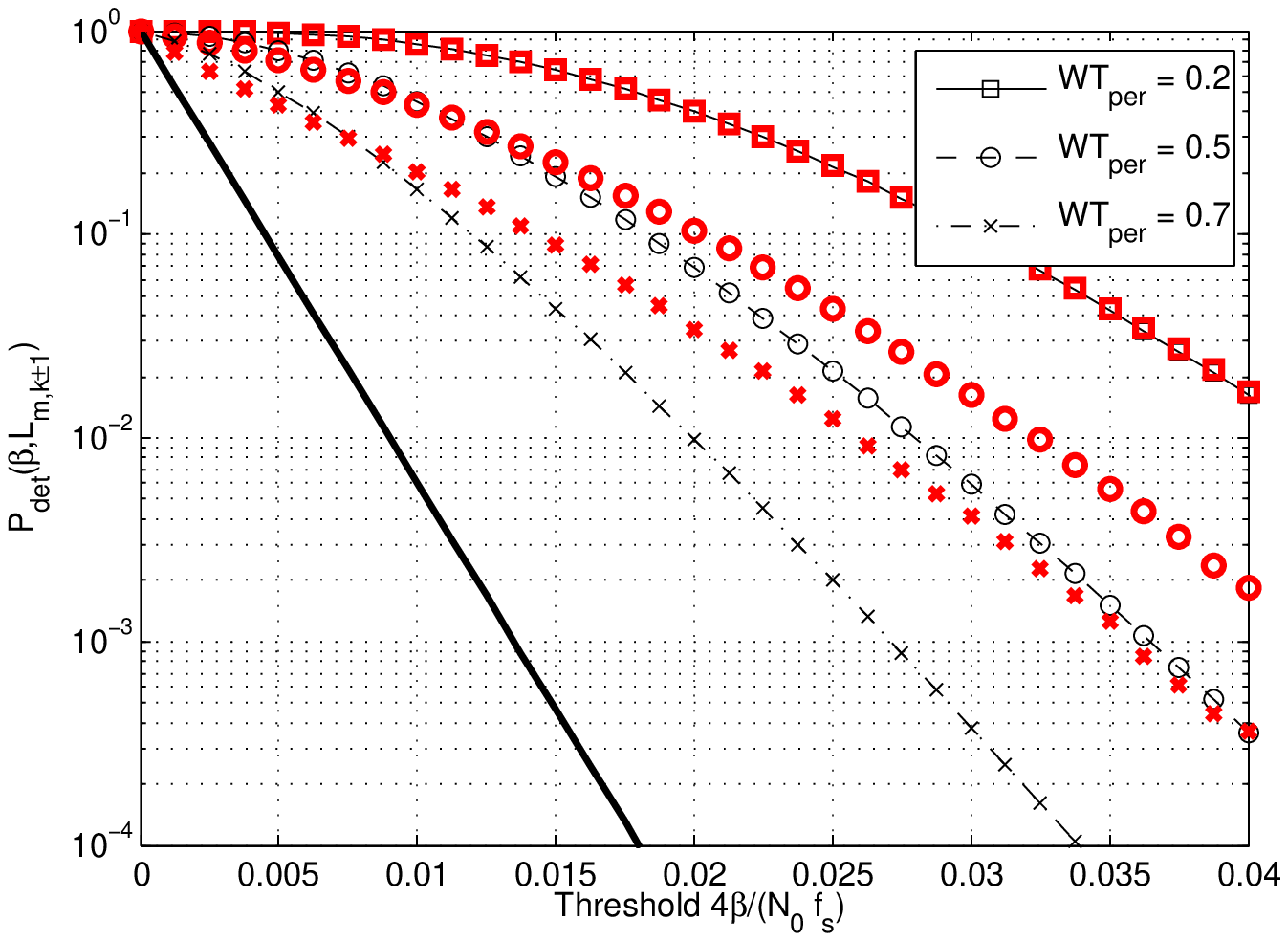}
 \caption{Cell detection probability $\pdetl{L_{m,k\pm1}}$ for different Doppler bin widths. Simulated (bold markers) and analytic (lines) results are shown for the Doppler bin adjacent to the correct one. The thick solid line indicates $\pdetl{0}=\pfab(\beta)$.}
 \label{fig:pdet_adj}
\end{figure}

\begin{figure}[t]
 \centering
 \includegraphics[width=0.47\textwidth]{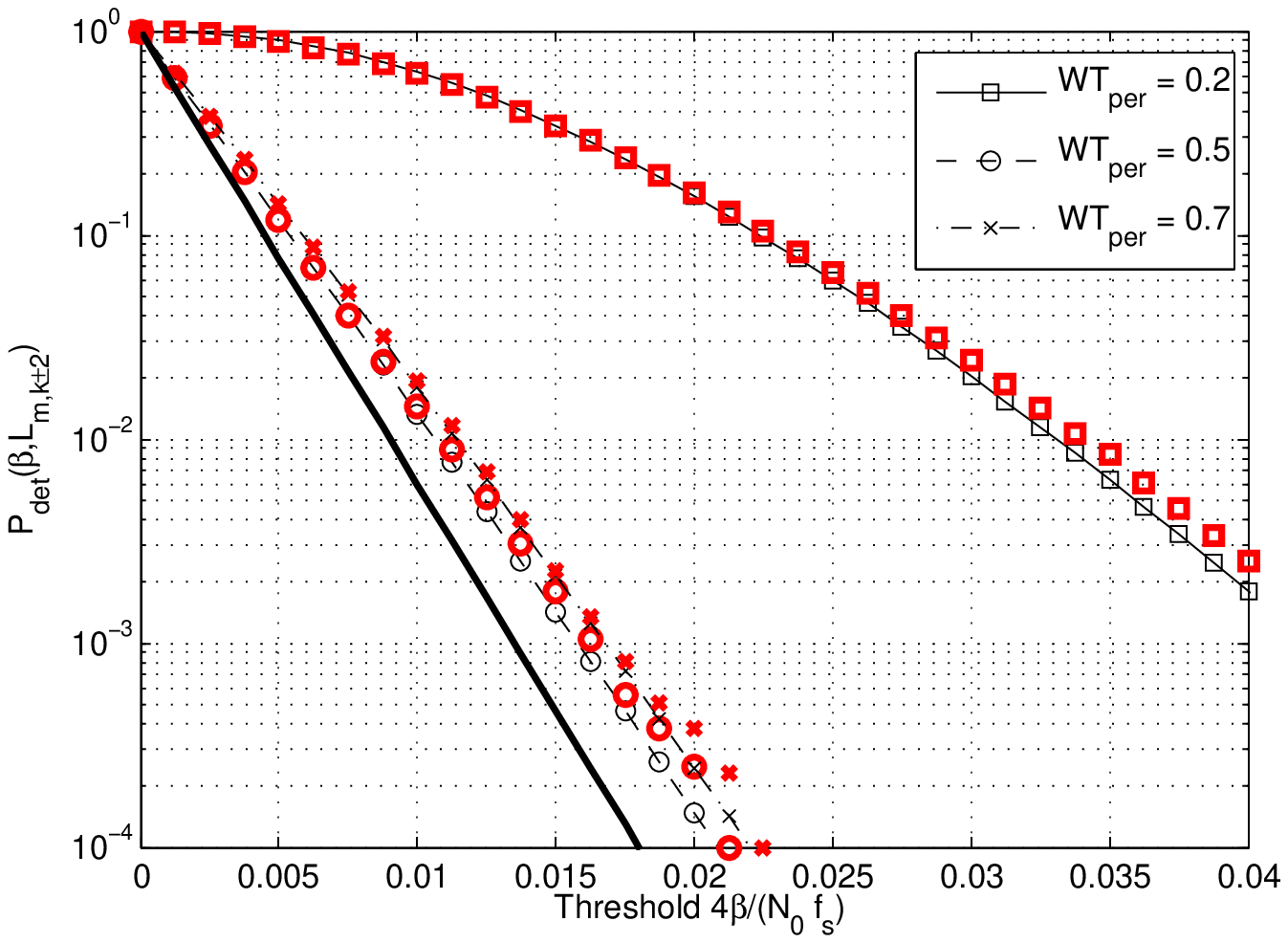}
 \caption{Cell detection probability $\pdetl{L_{m,k\pm2}}$ for different Doppler bin widths. Simulated (bold markers) and analytic (lines) results are shown for the Doppler bin second adjacent to the correct one. The thick solid line indicates $\pdetl{0}=\pfab(\beta)$.}
 \label{fig:pdet_sec}
\end{figure}

Figs.~\ref{fig:pdet_corr},~\ref{fig:pdet_adj}, and~\ref{fig:pdet_sec} show a comparison between the simulated and analytic cell detection probabilities for the correct, the first, and the second adjacent Doppler bins for different bin widths. \textcolor{\red}{As expected, in all three cases a larger relative Doppler bin width $\DW\tper$ leads to a decreased detection probability. Conversely, for a fixed bin width the probability decreases with distance to the correct bin, i.e., the detection probability is smaller in bins with index $k\pm2$ than in bins with index $k\pm1$.} In addition to that, it is shown (thick lines) that by not considering Doppler bin widths at all, the results would be too optimistic, leading to an overestimation of the global detection probability (see Section~\ref{sec:globalProbs}).

It can be seen that there is a good match between the analytic and the simulated results, except for the Doppler bin directly adjacent to the correct one. In this particular case, a separate analysis showed that for $\hat{k}=k\pm1$ and for larger Doppler bin widths,~\eqref{eq:approximate} is not a good approximation of~\eqref{eq:accurate}, which consequently leads to significant deviations. This is related to the fact that the linear approximation of the conditional PDF $f_{Y|L}(y,\lambda)$ is not sufficiently accurate for these choices of parameters (see Appendix~\ref{app:linear}).

Since the cell false alarm probability does not depend on the non-centrality parameter, it is not affected by the Doppler bin width.

\subsection{Influence of Doppler Bin Width on Global Probabilities}\label{sec:globalProbs}
\begin{figure}[t]
 \centering
 \includegraphics[width=0.47\textwidth]{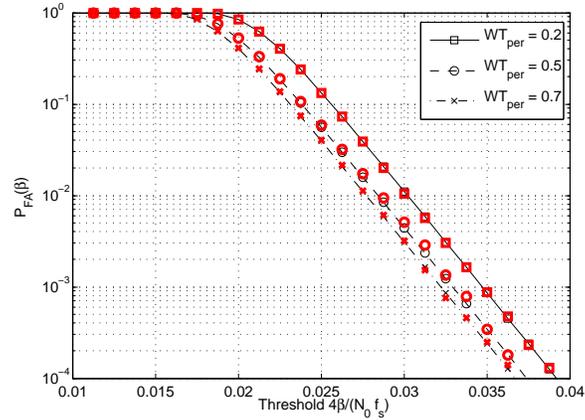}
 \caption{Global false alarm probability $\pFA$ for different Doppler bin widths. Simulated (bold markers) and analytic (lines) results are shown.}
 \label{fig:pfa}
\end{figure}

Combining these probabilities to global detection and false alarm probabilities shows another picture: Here, the effect of a greater number of noise cells becomes apparent, showing that smaller Doppler bins do not necessarily lead to improved performance. For example, Fig.~\ref{fig:pfa} shows that the global false alarm probability increases for smaller Doppler bins, i.e., for an increasing number $K$ of bins -- this is intuitively understood by looking at~\eqref{eq:PFA_Borio}. Note that in accordance with Section~\ref{sec:propabilities} a global false alarm occurs if a cell is detected in the absence of a signal. Therefore, the global false alarm probability is the same for the naive assumption as for the refined model presented in this work. 

\begin{figure}[t]
 \centering
 \includegraphics[width=0.47\textwidth]{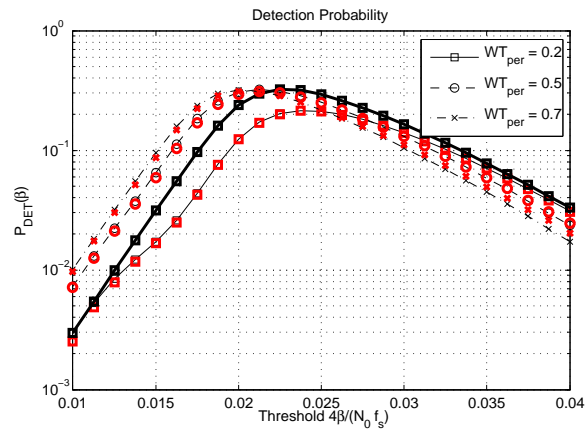}
 \caption{Global detection probability $\pDET$ for different Doppler bin widths. Simulated (bold markers) and analytic (lines) results are shown. \textcolor{\red}{The thick line represents the naive assumption of $L_{m,k}=L_{\max}$ and $L_{m,\hat{k}}=0$ for all $\hat{k}\neq k$ (shown only for $\DW\tper=0.2$).}}
 \label{fig:pdet}
\end{figure}

Fig.~\ref{fig:pdet} shows the probability of detecting the correct cell (i.e., only the correct Doppler bin is accepted, $M=0$) in a serial search which, especially for small Doppler bins and low thresholds, suffers from high false detection (i.e., stopping at a wrong cell in the case of a present signal) and false alarm rates. The additional bend in the curve for $\DW\tper=0.2$ near the maximum is due to significant energy of the adjacent Doppler bins, which increases the probability of triggering a false detection. As shown, low thresholds benefit from larger Doppler bins (small $K$, little energy in adjacent bins), whereas the opposite is true for larger values of $\beta$. There, small Doppler bins lead to a high $L_{m,k}$, i.e., a high cell detection probability for the correct cell, whereas false detections are unlikely due to the large threshold. To contrast the refined model against the naive assumption we also plotted the results of the latter for $\DW\tper=0.2$ in Fig.~\ref{fig:pdet}. As can be seen, the match is quite good for very high thresholds (where a false detection is unlikely) and for very low thresholds (where the false alarm probability is high). In the relevant area near the maximum, however, the difference between the models is significant, leading to an overestimation of the detection performance based on the naive assumption.

As Figs.~\ref{fig:pfa} and~\ref{fig:pdet} show, the analytic results are widely validated by the simulations, despite the fact that the detection probability for the Doppler bins with indices $k\pm1$ was underestimated by the theoretical approximation.

\subsection{Receiver Operating Characteristics}
\label{sec:ROC}
\begin{figure*}[t]
 \centering
 \includegraphics[width=0.7\textwidth]{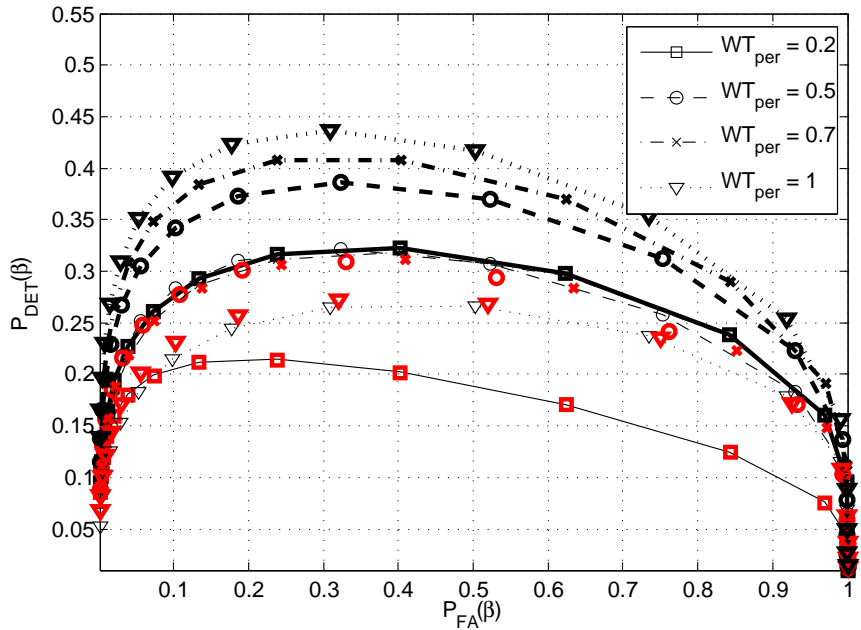}
 \caption{Receiver Operating Characteristics for $\frac{C}{N_0}=40$~dBHz. Simulated (bold markers) and analytic (lines) results are compared to the naive assumption of $L_{m,k}=L_{\max}$ and $L_{m,\hat{k}}=0$ for all $\hat{k}\neq k$ (thick lines)}
 \label{fig:ROC}
\end{figure*}

Fig.~\ref{fig:ROC} plots the global detection probability as a function of the global false alarm probability for $M=0$, i.e., for the scenario where only the correct Doppler bin is accepted. On the one hand, as it was expected, neither very small nor very large Doppler bins perform well: In the former case, the \textcolor{\red}{false detection probability due to signal energy in adjacent Doppler bins is increased}, while in the latter the detection probability is decreased. On the other hand, medium Doppler bin widths in the order of $\DW\tper=0.5$ to $\DW\tper=0.7$, which correspond roughly to the widely used values of 500~Hz to 666~Hz, turn out to perform optimally in terms of receiver operating characteristics, at least for this particular value of $\frac{C}{N_0}$. Thus, our analysis confirms the prevailing design practices~\cite{Borre_SDRGPS,Ward_GPSAcquisition,Kaplan_GPS,Tsui_GPSReceivers,Borio_PhD}, albeit they are justified by different arguments.

By neglecting the influence of Doppler bin widths, one again can see that the obtained results are overly optimistic. Looking at the thick lines in Fig.~\ref{fig:ROC}, which consider the number of bins, $K$, but not the influence on the non-centrality parameters $L_{m,\hat{k}}$, one is tempted to conclude that large Doppler bins outperform smaller ones. This result, however, is based entirely on the assumption that there is just one signal cell, and that the energy contained in this cell is independent of the Doppler bin width.

We conclude that the naive model suffers from significant inaccuracies by neglecting the influence of the Doppler bin width. The assumption of a single signal cell does not hold for small Doppler bins. It does hold, however, for large Doppler bins, and we suggest that the accuracy of the naive model can already be improved significantly by taking the influence of the Doppler bin width on the non-centrality parameter into account.

As Fig.~\ref{fig:ROC} shows, despite model inaccuracies pointed out in Section~\ref{sec:cellProbs} and in Appendix~\ref{app:linear}, our theoretical framework matches the simulations quite well. In addition to that, compared to the naive assumption of a single signal cell, our model leads to dramatically improved estimates of receiver operating characteristics.

\begin{figure}[t]
 \centering
 \includegraphics[width=0.47\textwidth]{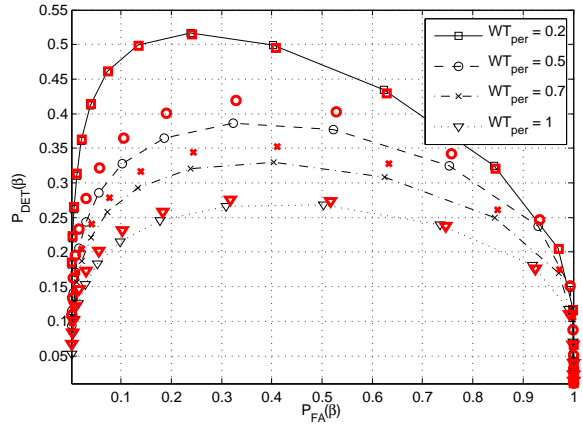}
 \caption{Receiver Operating Characteristics for $\frac{C}{N_0}=40$~dBHz. Successful detection occurs if the serial search is stopped either at the correct Doppler bin or at the two adjacent ones. Simulated (bold markers) and analytic (lines) results are compared.}
 \label{fig:ROCboth}
\end{figure}

Fig.~\ref{fig:ROCboth} shows the receiver operating characteristics for the scenario introduced in Section~\ref{ssec:adjacent}, i.e., where a search stopped at a Doppler index $\hat{k}\in\{k-M,\dots,k+M\}$ is successful (given the correct code phase is chosen). For simplicity, we allow here only a detection in Doppler bins immediately adjacent to the correct one, i.e., $M=1$. The receiver operating characteristics (ROCs) resulting from the naive assumption are not plotted in this case, since they do not change significantly compared to Fig.~\ref{fig:ROC}.

As it can be seen, smaller Doppler bins clearly outperform larger Doppler bins, since they benefit from increased cell detection probabilities in the Doppler bins adjacent to the correct one. It appears that this way the negative influence of an increased number of Doppler bins, as it is suggested by the naive assumption (cf.~Fig.~\ref{fig:ROC}), can be mitigated completely.

Comparing analytic results with simulations, one can see that especially for very large and very small Doppler bins a good match is achieved. In the former case this corresponds to the fact that the cell detection probability in the Doppler bin adjacent to the correct one is negligible, while in the latter the good match is due to the fact that the approximation of the non-centrality parameter holds well in this case (cf.~Fig.~\ref{fig:pdet_adj}).

For large relative Doppler bin widths, e.g., $\DW\tper=1$, the receiver operating characteristics seem to be unexpectedly bad. However, our analysis presents a lower bound on the performance by assuming a uniformly distributed Doppler frequency and by applying a serial search. More prior knowledge and better search methods would most likely lead to better results. Furthermore, our analysis does not take the verification stage of a typical acquisition process into account, which further improves the receiver operating characteristics by confirming or falsifying threshold crossings~\cite{Grewal_GPSIntertialNavigation}.

\subsection{A Real-World Scenario}
\begin{figure}[ht]
 \centering
 \includegraphics[width=0.47\textwidth]{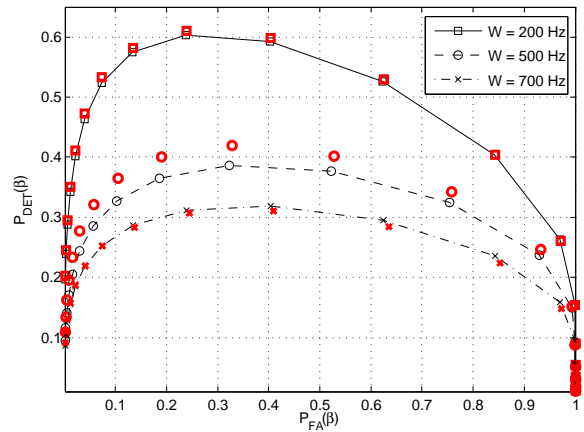}
 \caption{Receiver Operating Characteristics for $\frac{C}{N_0}=40$~dBHz. Successful detection is defined such that a tracking stage with a pull-in range of 500~Hz (see text) is likely to successfully utilize the acquisition results. Simulated (bold markers) and analytic (lines) results are compared.}
 \label{fig:ROC2adj}
\end{figure}

We assume a frequency-locked loop (FLL) of the subsequent tracking stage with a pull-in range of 500~Hz. Since the bandwidth of the FLL shall be related to the integration period $\tper=1$~ms of the acquisition stage~\cite{Tsui_GPSReceivers}, this is a realistic assumption. We now have to choose the Doppler bin widths and the number $M$ of acceptable adjacent bins such that the FLL can lock to the signal with high probability. In particular, for $\DW=700$~Hz only the correct Doppler bin ($M=0$) was considered as successful detection. For a bin width of 500~Hz the correct and the immediately adjacent bins lead to a detection, which limits the remaining Doppler difference to at most 750~Hz. However, as Fig.~\ref{fig:sinc} shows, a remaining Doppler difference larger than 500~Hz is unlikely due to the small non-centrality parameter. For the small bin widths of $\DW=200$~Hz two adjacent bins were accepted on either side of the correct bin, leading to a maximum remaining Doppler difference of 500~Hz.

As can be already expected by looking at Fig.~\ref{fig:ROCboth}, smaller Doppler bins will outperform larger ones in terms of receiver operating characteristics. This also holds in the practical example depicted in Fig.~\ref{fig:ROC2adj}, where an excellent performance is achieved with $\DW=200$~Hz. The only drawback of using smaller Doppler bins (which lead to a larger number of cells to search over) is the increased computational complexity and acquisition time. A closer analysis of this effect is within the scope of future work.

\section{Conclusion}
In this article, the influence of Doppler bin widths on GPS acquisition performance is analyzed. Analytic expressions, linking the Doppler bin width to the detection probabilities, are derived and evaluated. These expressions extend the conventionally used detection and false alarm probabilities, which assume that the search region is populated by noise-only cells except for a single signal cell. This assumption, specifically, is shown to be overly optimistic.

It is shown that small Doppler bins increase the probability of a global false alarm, but increase the detection probability within the correct Doppler bin. Since for small bins also the detection probability in adjacent bins is increased, the search strategy has to be adapted accordingly: If only the correct Doppler bin leads to successful detection, significant signal energy in adjacent bins has adverse effects on the receiver operating characteristics. Conversely, if in addition to the correct bin a certain number of adjacent Doppler bins is accepted, it turns out that small Doppler bins outperform larger ones in terms of receiver operating characteristics.

%

\appendices
\section{Derivation of Refined Detection Probabilities}
\label{app:derivation}
\newcommand{\pdb}[1]{P_{DB}(#1)}
\newcommand{\pdbn}[1]{\overline{P}_{DB}(#1)}
We now outline the derivation of~\eqref{eq:pcpsM} (i.e., the global detection probability for a serial search over all code phases for each Doppler bin) for $M=1$. The generalization to accepting more than one adjacent Doppler bin ($M>1$) and the derivation of~\eqref{eq:pfsM} follow along the same lines.

Assume that we are in the Doppler bin with index $\hat{k}$. We now serially search all code phases, starting from the first code phase until we reach the $N$-th code phase. The probability that we stop at the correct code phase is equal to the probability that the first code phase is correct and we stop there, plus the probability that the second code phase is correct and we stop there after not getting a false alarm at the first code phase, plus the probability that the third code phase is correct, etc. Mathematically, assuming that the correct code phase is uniformly distributed among the $N$ possible code phases,
\begin{multline}
  \frac{1}{N}\pdetl{L_{m,\hat{k}}}+\frac{1}{N}\pdetl{L_{m,\hat{k}}}\pfan(\beta)+\dots\\+\frac{1}{N}\pdetl{L_{m,\hat{k}}}\pfan^{N-1}(\beta)\\
=\pdetl{L_{m,\hat{k}}}\frac{1-\pfan^N(\beta)}{N\pfab(\beta)}=:\pdb{\hat{k}}.
\end{multline}
Let further $\pdbn{\hat{k}}=\pdetln{L_{m,\hat{k}}}\pfan^{N-1}(\beta)$ denote the probability that in the Doppler bin with index $\hat{k}$ no detection is triggered.

We now search over all $K$ Doppler bins: The probability that we stop at the correct code phase in the correct Doppler bin is equal to the probability that the first bin is correct and we stop at the correct phase, plus the probability that the second bin is correct and we do not stop at the first, but at the correct phase in the second bin, etc. Writing this down mathematically, assuming that all $K$ bins are correct with equal probability, we get
\begin{multline}
 \frac{1}{K}\pdb{k}+\frac{1}{K}\pdbn{k-1}\pdb{k}+\dots\\
\frac{1}{K}\pdbn{k-K+1}\dots\pdbn{k-1}\pdb{k}\\
=\frac{1}{K}\pdb{k}\sum_{n=0}^{K-1}\prod_{l=1}^n\pdbn{k-l}.
\end{multline}
where we used the convention that $\prod_{l=1}^0(\cdot)=1$. Note that this sum contains $K$ terms, since there are $K$ possibilities to stop the search at the correct Doppler bin.

The probability that we stop at the correct code phase in the Doppler bin coming \emph{before} the correct one is equal to the probability that the second bin is correct and we stop at the correct phase in the first, plus the probability that the third bin is correct and we do not stop at the first, but at the correct phase in the second bin, etc. In formulas,
\begin{multline}
 \frac{1}{K}\pdb{k-1}+\frac{1}{K}\pdbn{k-2}\pdb{k-1}+\dots\\
=\frac{1}{K}\pdb{k}\sum_{n=0}^{K-2}\prod_{l=1}^n\pdbn{k-l-1}.
\end{multline}
Since the first Doppler bin in this case cannot be the correct one (we cannot stop our search at the zeroth bin), this sum contains only $K-1$ terms. Along the same lines we obtain for the probability of stopping the search at the Doppler bin \emph{after} the correct one:
\begin{multline}
 \frac{1}{K}\pdbn{k}\pdb{k+1}\\+\frac{1}{K}\pdbn{k-1}\pdbn{k}\pdb{k+1}+\dots\\
=\frac{1}{K}\pdb{k+1}\sum_{n=1}^{K-1}\prod_{l=1}^n\pdbn{k-l+1}
\end{multline}
Since in this case the $K$-th bin cannot be the correct one (we assumed that the $(k+1)$-th bin triggered a detection), also this sum consists of $K-1$ terms. The global detection probability for $M=1$ now equals the sum of these three probabilities, and we obtain
\begin{equation}
 \pDET=\frac{1}{K}\sum_{q=-1}^1\pdb{k+q}\sum_{n=n'}^{K'-1}\prod_{l=1}^n\pdbn{k-l+q}
\end{equation}
where $K'=\min\{K,K+q\}$ and $n'=\max\{0,q\}$. If we now insert the equations for $\pdbn{\hat{k}}$ and $\pdb{\hat{k}}$ we obtain~\eqref{eq:pcpsM} for $M=1$.

\section{Linear Approximation of the Conditional PDF}
\label{app:linear}
\begin{figure}[ht]
 \centering
 \includegraphics[width=0.47\textwidth]{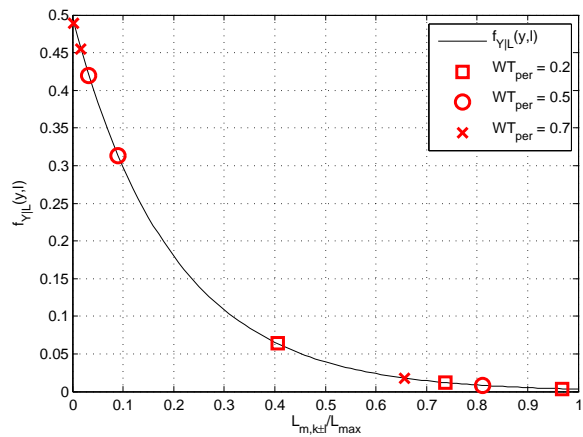}
 \caption{Conditional probability $f_{Y|L}(y,l)$ evaluated $\frac{C}{N_0}=40$~dBHz and $y=0.0225$. Markers indicate Doppler bin boundaries for designated relative bin widths $\DW\tper$.}
 \label{fig:NonCent}
\end{figure}

In this appendix, we show that under a linearity assumption~\eqref{eq:accurate} can be approximated by~\eqref{eq:approximate}. Thus, let us first assume that
\begin{equation}
 f_{Y|L}(y,\lambda) \approx k(\lambda-\expec{L})+f_{Y|L}(y,\expec{L}).
\end{equation}
Substituting this into~\eqref{eq:accurate} leads to
\begin{IEEEeqnarray}{RCL}
 \pdetl{L} &=& \int_{\beta}^{\infty} \int_{-\infty}^{\infty}  f_{Y|L}(y,\lambda) f_L(\lambda) d\lambda dy\\
 &\approx& \int_{\beta}^{\infty}f_{Y|L}(y,\expec{L})dy\notag\\&&+ \int_{\beta}^{\infty}\int_{-\infty}^{\infty} k(\lambda-\expec{L}) f_L(\lambda)d\lambda dy\\
&=& \int_{\beta}^{\infty}f_{Y|L}(y,\expec{L})dy,
\end{IEEEeqnarray}
which is~\eqref{eq:approximate}. In Fig.~\ref{fig:NonCent} this linearity assumption is illustrated. The range between the rightmost marker and the value $L_{m,k\pm l}/L_{\max}=1$ represents the correct Doppler bin. The range between the two rightmost (leftmost) markers indicates the first (second) adjacent Doppler bin ($k\pm1$ and $k\pm2$, respectively; compare to Fig.~\ref{fig:sinc}). As it can be seen, both the range between the two leftmost markers and the range between the rightmost marker and the boundary can be well approximated by a line. Thus,~\eqref{eq:approximate} approximates~\eqref{eq:accurate} well and the corresponding detection probabilities are matching simulations (see Figs.~\ref{fig:pdet_corr} and~\ref{fig:pdet_sec}). The range between the two leftmost markers, on the other hand, is only approximately linear for $\DW\tper=0.2$. For bin widths $\DW\tper=0.5$ and $\DW\tper=0.7$ the linearity assumption clearly does not hold. As a consequence, the detection probability obtained for the adjacent Doppler bin is valid only for $\DW\tper=0.2$ (see Fig.~\ref{fig:pdet_adj}).

\section*{Acknowledgments}
The authors would like to thank Daniel Arnitz for valuable suggestions and feel particularly indebted to the anonymous reviewers helping to improve the manuscript. This work was partially funded by the Austrian Research Promotion Agency under the project ``SoftGNSS 2'', project number 819682.

\bibliographystyle{IEEEtran} 
\bibliography{IEEEabrv.bib,/afs/spsc.tugraz.at/project/GNSS/1_d/Papers/GPS_Papers.bib,/afs/spsc.tugraz.at/user/bgeiger/includes/textbooks.bib,/afs/spsc.tugraz.at/user/bgeiger/includes/myOwn.bib}

\end{document}